# A Phenomenological Cost Model for High Energy Particle Accelerators [1]

Vladimir Shiltsev

*Fermi National Accelerator Laboratory, PO Box 500, Batavia, IL 60510, USA*

*Abstract*

Accelerator-based high-energy physics have been in the forefront of scientific discoveries for more than half a century. The accelerator technology of the colliders has progressed immensely, while the beam energy, luminosity, facility size, and cost have grown by several orders of magnitude. The method of colliding beams has not fully exhausted its potential but has slowed down considerably in its progress. In this paper we derive a simple scaling model for the cost of large accelerators and colliding beam facilities based on costs of 17 big facilities which have been either built or carefully estimated. Although this approach cannot replace an actual cost estimate based on an engineering design, this parameterization is to indicate a somewhat realistic cost range for consideration of what future frontier accelerator facilities might be fiscally realizable.

**Introduction**

The forecast for the far future of colliders beyond say, the 2030's, requires an understanding of several factors: scientific goals and desired energy reach, the possible available resources, and a glimpse of suitable accelerator technology of the future. Affordable cost of the frontier facility is crucial. As of today, the world's particle physics research budget can be estimated to be roughly 3B$ per year. Under the assumption that such financial situation will not change by much in the future and that not more than 1/3 of the total budget can be dedicated to construction of the next energy frontier collider over approximately a decade, one can estimate the cost of a globally affordable future facility to be about or less then 10B$ (in current prices.) The desired features of such a flagship machine are [1]: ideally, its energy reach should significantly ($> \times 10$) exceed that of the LHC (14 TeV center-of-mass energy), it should not exceed ~10 km in length, and its AC wall power consumption should stay under ~100MW. To understand whether there are opportunities for building such an accelerator within the above-mentioned financial constraints, an analysis of the known costs of large accelerator facilities is needed.

---

[1] shiltsev@fnal.gov

**Analysis of Known Costs of Large Accelerator Facilities**

There are several factors which need to be taken into account while discussing the costs of 17 large accelerators of the past two decades [2]: a) firstly, some of the facilities had been seriously considered for construction and well estimated but not built, e.g., the Superconducting Super Collider (SSC) in Texas, the Very Large Hadron Collider (VLHC) in Illinois, SLAC's proposal for the Next Linear Collider (NLC), linear $e+e-$ collder TESLA at DESY; others are still being under consideration – the International Linear Collider (ILC), Project X at Fermilab, CERN's Super Proton Linac (SPL), Compact Linear Collider (CLIC), the Beta-Beam and Neutrino Factory projects; and there are several which have been built or are being built - Fermilab's Main Injector proton synchrotron (MI), Brookhaven's Relativistic Heavy Ions Collider (RHIC), Spallation Neutron Source (SNS) at ORNL, CERN's Large Hadron Collider (LHC), European Spallation Source (ESS) at Lund and X-ray Free Electron Laser (XFEL) at DESY, and Facility for Antiproton and Ion Research (FAIR) at Darmstadt. b) secondly, the scope of the facility construction significantly varied and many facilities heavily piggybacked on either existing infrastructure or tunnel or injection complex – RHIC and LHC are the most notable examples – while others were/are true "green field" facilities, c) thirdly, while almost all accelerator facilities have in common such components as accelerator tunnels and high power electrical power, they differ significantly in the major accelerator technologies they are built upon – such as normal-conducting (warm) or superconducting magnets or RF linacs. The most important difference for the above listed projects is often the methodology of the cost estimates. Some construction projects come out with the cost estimates of the industrial contracts for major items like civil engineering, the accelerator elements and corresponding labor requirements (such approach is often referred as "European accounting") while others do literally full accounting of all associated expenses ("US accounting") E.g., all scientific facilities supported by the US DOE Office of Science are required to prepare and report estimates of "the total project cost" (TPC) that include not only the cost of the technical components and conventional systems and associated labor, but also costs of the required R&D, development of the engineering design, program management, escalation, contingency, overhead funds, project-specific facility site development, sometimes – detectors, etc. The difference between the TPC and the "European accounting" was the subject of detail investigations during discussions on the US participation/contribution to the TESLA and ILC linear colliders projects and the International Thermonuclear Experimental Reactor (ITER) and was estimated to be in the range (factor of) 2.0-2.5 [2]. Finally, one can also add that even the TPC estimates for not-yet-built facilities should be understood only in a probabilistic sense as, for example, the contract bidding process in not fully predictable [3], and, moreover, "…costing a new project in advanced technology is not just an accounting task, but rather an attempt to organize complexity and uncertainty" [4].

---

[2] though many accelerators have been built in the past and their construction costs are known, we limit ourselves only to the machines which are most relevant to our discussion of the mid-term and long-term future of high energy physics – namely, those built or designed relatively recently so their accelerator technologies are not too obsolete now due the progress in the field and whose construction costs can be scaled to nowadays using standard inflation indices with reasonable degree of accuracy.

Table 1 summarizes an attempt to present the cost ranges - sometimes as large as ±35% - for the 17 recent large accelerator projects in the same TPC-methodology and in "present day" dollars, i.e., taking into account the corresponding inflation and currency conversion indices [5]. The Table convincingly shows that in general, the cost of the facility is higher if the beam energy is higher, but at the same time it also hints that the cost per TeV (or GeV) depends on the technology of choice and that facilities that require higher beam and site power tend to be more expensive.

Table 1: Cost estimates of large accelerator facilities. Energy is center-of-mass energy for colliders and maximum beam energy for single beam machines. TPC stands for "Total Project Cost" (see text); technology types – "SC/NC Mag" for superconducting/normal-conducting magnets, "SC/NC RF" – for superconducting/normal-conducting magnets. Total length numbers include injector complex tunnels.

|  | Cost (B$) Year | Energy (TeV) | Accelerator technology | Comments | Length (km) | Site power (MW) | TPC range (Y14 B$) |
|---|---|---|---|---|---|---|---|
| SSC | 11.8B$ (1993) | 40 | SC Mag | Estimates changed many times [6, 7, 8] | 87 | ~100 | 19-25 |
| FNAL MI | 260M$ (1994) | 0.12 | NC Mag | "old rules", no OH, existing injector [9] | 3.3 | ~20 | 0.4-0.54 |
| RHIC | 660M$ (1999) | 0.5 | SC Mag | Tunnel, some infrastructure, injector re-used [10] | 3.8 | ~40 | 0.8-1.2 |
| TESLA | 3.14B€ (2000) | 0.5 | SC RF | "European accounting" [11] | 39 | ~130 | 11-14 |
| VLHC-I | 4.1B$ (2001) | 40 | SC Mag | "European accounting", existing injector [12] | 233 | ~60 | 10-18 |
| NLC | ~7.5B$ (2001) | 1 | NC RF | ~6B$ for 0.5 TeV collider, [13] | 30 | 250 | 9-15 |
| SNS | 1.4B$ (2006) | 0.001 | SC RF | [14] | 0.4 | 20 | 1.6-1.7 |
| LHC | 6.5 BCHF (2009) | 14 | SC Mag | existing tunnel, injector & infrstr., no OH, R&D [15] | 27 | ~40 | 7-11 |
| CLIC | 7.4-8.3B CHF(2012) | 0.5 | NC RF | "European accounting" [16] | 26 | 250 | 12-18 |
| Project X | 1.5B$ | 0.008 | SC RF | [17] | 0.4 | 37 | 1.2-1.8 |

| | | | | | | |
|---|---|---|---|---|---|---|
| | (2009) | | | | | | |
| XFEL | 1.2B€ (2012) | 0.014 | SC RF | in 2005 prices, "European accounting" [18] | 3.4 | ~10 | 2.9-4.0 |
| NuFactory | 4.7-6.5 B€(2012) | 0.012 | NC RF | Mixed accounting, w. contingency [19] | 6 | ~90 | 7-11 |
| Beta-Beam | 1.4-2.3B€ (2012) | 0.1 | SC RF | Mixed accounting, w. contingency [19] | 9.5 | ~30 | 3.7-5.4 |
| SPL | 1.2-1.6B€ (2012) | 0.005 | SC RF | Mixed accounting, w. contingency [19] | 0.6 | ~70 | 2.6-4.6 |
| FAIR | 1.2 B€ (2012) | 0.003-.08 | SC Mag | "European accountting" [20], 6 rings | ~6 | ~30 | 1.8-3.0 |
| ILC | 7.8 B$ (2013) | 0.5 | SC RF | "European accounting" [21] | 34 | 230 | 13-19 |
| ESS | 1.84B€ (2013) | 0.0025 | SC RF | "European accounting" [22,23] | 0.4 | 37 | 2.5-3.8 |

These observations lead to the following approach: one can de-compose the cost of each accelerator into only three parts which in aggregate total the known TPC (see Table 1.) Each part can be parameterized by a single, most relevant parameter of the facility. Namely, one major component of the TPC is the cost of civil engineering and construction (tunnels, surface buildings, connecting shafts, halls for experimental detectors, injection beamlines, beam dump halls, civil construction for injectors, tunnel infrastructure, etc.) and it will be parameterized vs the total tunnel length *L*; the cost of the accelerator components (RF cavities/modules, cryostats, input couplers, all magnets, beam instrumentation and feedback systems, vacuum systems, RF systems, injection and ejection, beam delivery systems for colliders, beam collimation and dumps, etc.) can be studied as a function of the center-of-mass energy *E* (or beam energy for single beam facilities); and finally, the cost of the facility infrastructure (electric power feeders and stations, cable trays, power distribution, main power connection, cryoplants and cryogenic distribution system, cooling and ventilation systems, safety systems, auxiliary systems, control systems, klystrons, magnet supplies, etc.) will be analyzed vs the total site electric power *P*. Refs. [3, 6-23] do provide sufficient information on the cost breakdown of the TPC among these three parts.

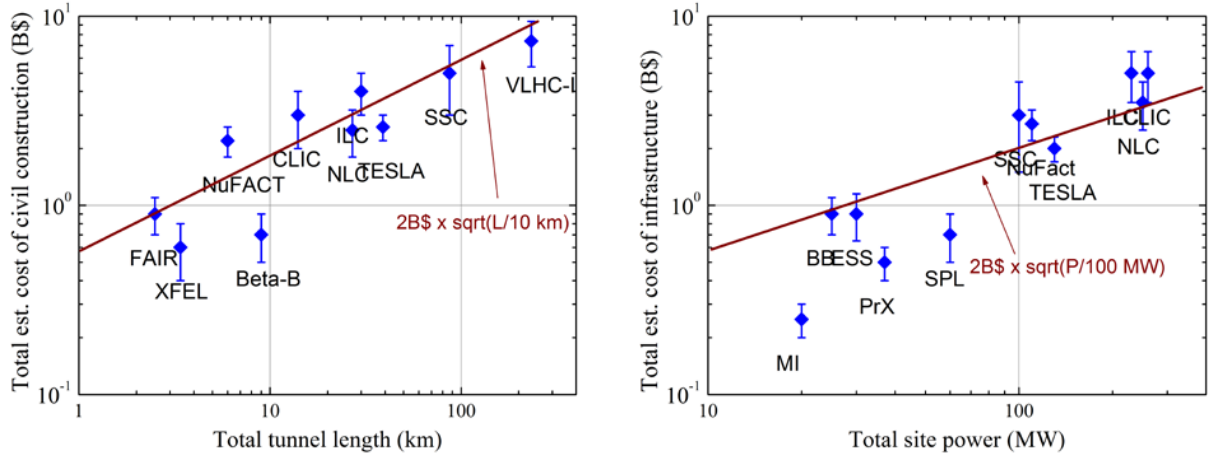

FIG. 1: a) left – total estimated cost of the civil construction for accelerator facilities vs cumulative length of their tunnels; b) right – total estimated infrastructure cost of the accelerator facilities vs their electric power consumption.

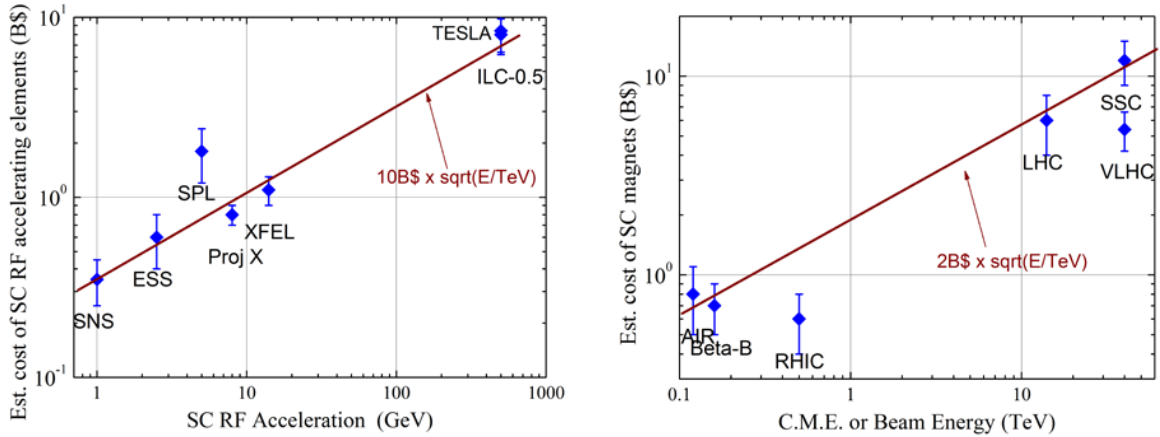

FIG. 2: a) left - estimated cost of the accelerating elements for accelerator facilities based on SC RF vs total beam acceleration; b) right – estimated cost of the SC magnets and associated elements for vs collider center of mass energy / single beam energy.

Figures 1 and 2 present the summary plots of such analysis. As shown in Fig.1a, the civil construction-related share of the TPC (not surprisingly) on average grows with the length of the tunnel. The error bars reflect uncertainties in both the TPC and its breakdown among the three main parts. The least square fit to the power function $y=aL^b$ gives coefficients $b=0.55\pm0.12$ and $a=1.1\pm0.4$ (if the length L in the units of 10 km and the cost in B$.) Significant dispersion in the parameters reflects obvious spread of the costs due to specific features of the accelerator tunnels and variation of other parameters, e.g., tunnel diameter, etc. With all that, it is quite remarkable that within about ±30% rather simple scaling of the civil construction cost of $2B\$\times(L/10km)^{1/2}$ describes the data – see straight solid line in Fig.1a. Fig.1b shows that similar type of scaling law can be used to describe the cost of the power infrastructure-related parts of the TPC, namely $\approx 2B\$\times(P/100MW)^{1/2}$ – with approximately the same fractional accuracy (the actual power law fit parameters are $b=1.0\pm0.2$ and $a=1.7\pm1.1$ (if the site power in the units of 100 MW and the cost in B$.) Much bigger cost differences are associated with accelerator technologies. For example, the estimated cost of the accelerating elements for accelerator facilities based on SC RF is on average five times that of the SC magnet based facilities - see Figs. 2a and 2b - the former can be approximated by $10B\$\times(E/TeV)^{1/2}$ (see the solid strait line in the Figure, the power law fit with $b=0.53\pm0.04$ and $a=9.1\pm1.4$) while the latter is close to some $2B\$\times(E/TeV)^{1/2}$ (the power law fit with $b=0.46\pm0.08$ and $a=1.2\pm0.27$.) There are only a few facilities in our Table 1 which employ normal-conducting RF structures or normal-conducting magnets, but following the same functional trend their shares of the cost of accelerator components can be approximated as $8B\$\times(E/TeV)^{1/2}$ and $1B\$\times(E/TeV)^{1/2}$ correspondingly.

Recombining all the costs together into the total cost one can now approximate $TPC\approx\Sigma\ a_i\ X^{b_i}_i$ where $X=(L, E, P)$ and $b_i$ and $a_i$ are the fit coefficients. Without losing much accuracy, the formulae can be further simplified by using the same exponent of $b=0.5$ – the square root dependence – for all three parts, leading to following phenomenological cost model for "green field" large accelerator facilities

$$TPC = \alpha\left(\frac{L}{10\ km}\right)^{1/2} + \beta\left(\frac{E}{1\ TeV}\right)^{1/2} + \gamma\left(\frac{P}{100MW}\right)^{1/2} \quad (1)$$

where, L is the total length of the tunnel, E is cm energy, P is total site AC power for the facility and $\alpha$, $\beta$, $\gamma$ are three coefficients, of which $\beta$ is the only one that varies for different accelerator technologies.

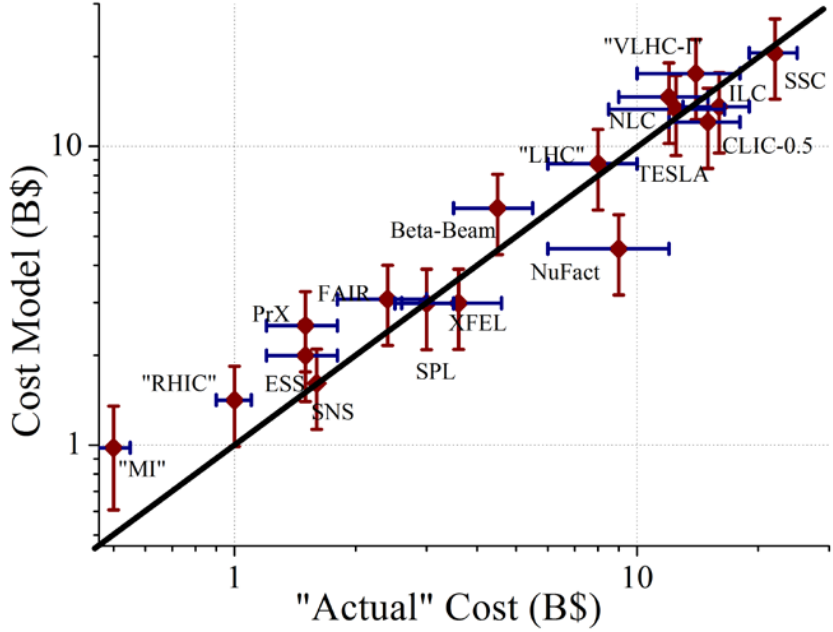

FIG. 3: The phenomenological cost model Eq. (1) vs actual cost for large accelerator facilities.

Figure 3 shows that within approximately ±30% the $\alpha\beta\gamma$-model matches the actual TPCs of all the facilities which we considered in the Table 1 above, with the coefficients $\alpha=\gamma=2B\$$ and technology dependent coefficient $\beta$ is equal *1B\$* for NC magnets, *2B\$* for SC magnets, *8 B\$* for NC RF and *10 B\$* for SC RF. Straight black solid line in Fig.3 indicates 1:1 correspondence; horizontal error bars show the TPC range from Table 1, vertical error bars reflect ±30% deviations from Eq.(1) (i.e., the range is about factor of 2 from minimum to maximum). Four facilities – MI, RHIC, LHC and VLHC-I - are emphasized by quotes "…", indicating that their known cost estimates do not account for significant "already existing" elements (infrastructure, tunnel or injector complex) – and that possibly explains why they are systematically above the 1:1 line in Fig.3 (i.e., the actual known cost is lower than the $\alpha\beta\gamma$-model TPC estimate of the "green field" facility Eq.(1).)

**Rough Cost Estimates for Possible Future Facilities**

Taking into account all the caveats of the phenomenological Eq.(1) – that it is for the US-accounting of the TPC of accelerator projects built from scratch, in the area where no previous



facilities exist, and that it is good at best within 30%, or for range of estimates of about a factor of 2 – one can apply the $\alpha\beta\gamma$-model to the future facilities employing the technologies similar to the ones of the past (but not necessarily exactly the same to them as the model seemingly reflects the progress in technologies made during the scale up of the past facilities to higher energies.) For example, the proposed TLEP $e+e-$ Higgs factory [24] would require $L$=80 km tunnel, $P\approx$300MW of site power, some $E_1$=6 GeV of CW SC RF system and normal-conducting magnet for the $E$=240 GeV c.o.m collider. Correspondingly, the combination of the costs in the $\alpha\beta\gamma$-model results in $TPC=2\times(80/10)^{1/2}+10\times(6GeV/1TeV)^{1/2}+1\times(250GeV/1TeV)^{1/2}+2\times(300/100)^{1/2}=$ *10.4B$ ±3B$* - that is in a decent agreement with a rough estimate of 5.5-7.5 BCHF under "European accounting" made in Ref.[24].

The high-energy CLIC $e+e-$ linear collider [25] with some $L$=60 km of tunnels, $P\approx$560MW of required site power, $E$=3 TeV of normal-conducting RF acceleration would cost about $TPC=2\times(60/10)^{1/2}+8\times(3TeV/1TeV)^{1/2}+2\times(560/100)^{1/2}=$ *23.5B$ ±7B$*. Even higher energy 6TeV $\mu+\mu-$ collider [26, 27] would require some $L$=20 km of tunnels, $P\approx$230MW of electric power, about $E_1$=50 GeV of CW SC RF system in the recirculating linear accelerator (RLA) and SC magnets in the collider ring. All that can be estimated to cost $TPC=2\times(20/10)^{1/2}+10\times(50\,GeV/1TeV)^{1/2}+2\times(6\,TeV/1\,TeV)^{1/2}+2\times(230/100)^{1/2}=12.9B\$ \pm 4B\$$ - i.e., almost a factor of four less expensive per TeV than CLIC. Notably, existence of the SPL-type proton linac would allow to shave additionally some 3-4 B$ off the TPC of the muon collider facility. Of course, there are caveats associated with a muon collider as it is first-of-a-kind operating with unstable particles and significant R&D is still required to demonstrate its technical and performance feasibility [28].

Recently proposed 100 TeV proton-proton collider FCC [29, 30] will employ SC magnets, and will require 100 km tunnel and some 400MW of site power. Therefore, the cost of such facility can be estimated as $TPC=2\times(100/10)^{1/2}+2\times(100\,TeV/1TeV)^{1/2}+2\times(400/100)^{1/2}=30.3B\$ \pm 9B\$$. Again, if some existing accelerators can be re-used as a part of the needed 5-7 TeV injector, then some 6-10 B$ can potentially be shaved off the TPC. In comparison, the LHC energy upgrade [31] would need some 20 T SC magnets for a 33 TeV $p$-$p$ collider in the existing LEP tunnel and additional ~100 MW of electric power, therefore, the cost of such upgrade is expected to be about *2×(33 TeV /1 TeV)$^{1/2}$+2×(100/100)$^{1/2}$=4.8B$ ±1.5B$*.

In some cases, one may apply the $\alpha\beta\gamma$-model to the future facilities which will be based on new, not-yet-fully-tested acceleration methods but heavily employ the existing technologies. The



Argonne Flexible Collider (AFC) concept of a 3 TeV *e+e-* machine [32] is based on dielectric-wakefield acceleration (DWA) principle in which electric fields in the diamond structures are excited by low energy high power electron bunches produced by traditional means – namely, in the 0.86 GeV pulsed normal conducting linacs. Twenty of such drive-beam linacs are needed for the AFC, which will also need some 20 km of tunnels (for the high energy linacs, beam delivery system and damping rings) and about 430MW of site power. Estimating the costs of these components gives us the lower limit of the total cost of $2\times(20/10)^{1/2}+20\times 8\times(0.86GeV/1TeV)^{1/2}+2\times(430/100)^{1/2} =21.7B\$ \pm 7B\$$. Of course, the cost of the DWA accelerating structures is not known yet, while it should be added to the TPC.

Similarly, it is not possible now to estimate the cost of the main (high energy) beam acceleration components in another high energy *e+e-* collider proposal based on the beam-plasma-wakefield acceleration (BPWA) [33], but one can at least estimate the cost of its 25 GeV ultra-high power SCRF drive-beam complex, civil construction for 20 km of tunnels for a 10 TeV facility and 540 MW site power infrastructure as $2\times(20/10)^{1/2}+10\times(25GeV/1TeV)^{1/2}+2\times(540/100)^{1/2} =9B\$ \pm 3B\$$. Just for reference, on average for the 17 facilities listed in Table 1, the cost of the main accelerating components (magnets, cavities, etc.) is about equal to the rest of TPC (infrastructure, civil construction, injectors, etc.).

**Discussion**

It is absolutely clear that the proposed phenomenological *αβγ*-model Eq.(1) cannot substitute actual cost estimates based on detailed facility-specific considerations, such as usually provided in what is called *Technical Design Reports*. On the other hand, there seems to be enough ground to believe in the validity of its application for estimates of the *cost range* of the TPC of large accelerator facilities. Significant deviation of other estimates for any future facility from the *αβγ*-model predication would definitely require an explanation of the difference compared to the model based on the data from 17 other machines.

The functional form of the power-law scaling is actually not that unusual. For example, the analysis of the costs of 270 tunnels worldwide (including industrial, subway, railway, etc.) [34] indicates that, though being dependent on the type of rock excavation and application, on average the cost scales as $L^c D^d$, where $L$ and $D$ are the tunnel length and diameter and exponents are in the range $c=0.4-1, d=0.6-1.5$. Similar sets of data for other 100 tunnels in UK, Europe, Australia and Amaericas [35] can be fitted by $c≈0.5$ (i.e., as in our model.) The power-law cost scaling is typical in the electrical industry, e.g., the cost of the power transformers scales as its MVA rating in the power of about 0.5 [36], and it has physics explanation as the MVA capability scales as power of size of the transformer magnetic core (one of the biggest cost drivers.)

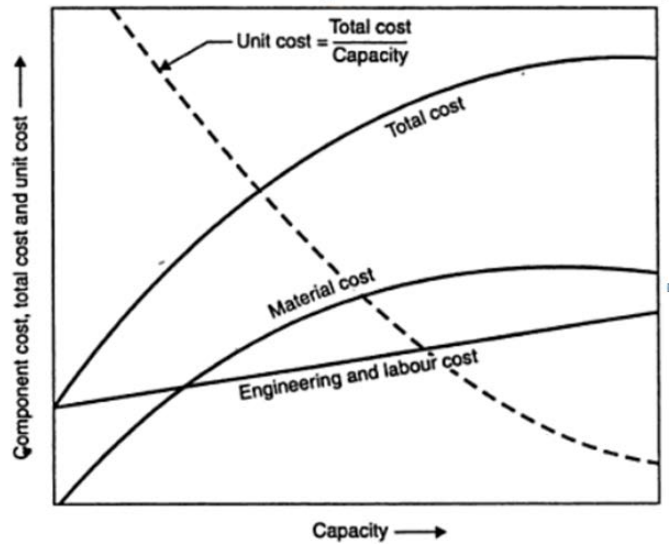

FIG.4: variation of the costs of power plant versus its capacity [37].

Similar dependencies are known in the power plan industry – see Fig.4 from [37] – the labor and engineering cost has significant intercept and increases slightly with the capacity of the unit, the slope of the material cost curve decreases slightly with increase of the capacity of the plant, as the result "…the total cost curve shows significant positive intercept at zero capacity which represents the cost of just maintaining an organization of men and plan ready to produce", and of course, the cost per unit of capacity goes down for larger plants.

Very similar arguments are valid for the TPC of an accelerator where "zero-acceleration" cost (the positive intercept) is usually a quite substantial part of the cost as it accounts for management, research and development, design work, cost of injection complex, and many other things which have to be addressed in addition to the construction of the accelerator proper. As it can be clearly seen from the "total cost" curve in Fig.4, the choice of *square root* fit is, therefore, more than natural as it allows for matching the reality over a wide range of the parameter (length, energy, and power.)

We should also note that the issue of the performance of the accelerator facility (e.g., luminosity of a collider) is not directly coupled to the facility parameters we used for our cost estimates. Indeed, even if the *L, E, P* (and cost) are defined, there might be ~order(s) of magnitude uncertainties related to important details such as beam quality. Attainment of the design or the ultimate performance of the facility can take quite substantial time [38].

Besides the financial feasibility, one should take into account the availability of experts. A simple "principle" Europe which is based on statistics of construction projects in Japan and widely accepted in the accelerator community claims that "one accelerator expert can spend intelligently 1 M$ in one year" [39]. As an illustration, the ILC TDR [21] estimates that some 13,000 man-years (FTEs, or full-time-equivalent) of accelerator scientists, engineers and technicians are needed over some 8 years of construction of the International Linear Collider. So, on average 1,600 trained people will be needed for installation, integration, testing and quality assurance, commissioning and all other related activities associated with 7.8B$ worth of materials and services budget – that is about 0.6M$ per person per year. Despite the lack of a crisp definition of who should be considered "an accelerator expert", we can estimate that the world-wide community of accelerator physicists and experienced engineers does not exceed 1200-1500 people and the total accelerator personnel (all scientists, engineers, technicians, drafters, etc) is about 4,000-4,500. Therefore, any plans for a really big facility at the scale of few B$ to 10B$ should take into account that significant time will be needed to get the required number of the people together.

Finally, coming back to the aspiration of a flagship machine with 10× the energy reach of the LHC – say, 100 TeV center-of-mass energy, which should not exceed ~10 km in length, and its AC wall power consumption should stay under $O(100MW)$, one can use the $\alpha\beta\gamma$-model to understand how economical should be the technology of choice, i.e. the technology cost coefficient $\beta$: if technologies of tunneling and electric power distribution will not change by much , i.e. if *α=γ=2B$,* then *β≈0.6B$,* or about twice cheaper than that of the normal-conducting magnets. While the technologies capable of delivering the required average accelerating gradient of 100TeV/10km=10 GeV/m do exist in principle – those are the acceleration by wakefields in plasma and in crystals [1] – their feasibility has not been demonstrated yet and their current costs are factors of at least 30 to 100 above the desired value of *β≈0.6B$/sqrt(TeV).* Only a comprehensive

R&D program can provide the answer to the question whether financially feasible accelerator based facilities can provide another order of magnitude step for the high energy physics frontier.

**Summary**

We have reviewed publicly available costs for 17 large accelerators of the past, present and those currently in the planning stage and attempted to reduce them to one methodology known as "the total project cost (TPC) or "the US accounting". The costs have been the broken up into three major parts corresponding to "civil construction", "accelerator components", "site power infrastructure" in such manner that they total the derived TPC ranges. We successfully attempted to parameterize the three cost component by just three parameters – the length of the tunnels $L$, the center-of-mass or beam energy $E$, and the total required site power $P$ and found that over almost 3 orders of magnitude of $L$, 4.5 orders of magnitude of $E$ and more than 2 orders of magnitude of $P$ the following cost model works with ~30% accuracy : *Total Project Cost* $\approx \alpha \times Length^{1/2}$ + $\beta \times Energy^{1/2} + \gamma \times Power^{1/2}$ where coefficients $\alpha$, $\gamma$ and accelerator technology dependent coefficient $\beta$ are defined in the text. The $\alpha\beta\gamma$-model has been applied to several proposed collider facilities and we obtained either their TPC ranges or the cost of their parts which are expected to be built on the base of the currently known accelerator technologies. We remark that besides the feasibility of the cost, very important are the feasibility of the performance and availability of expertise for large machine construction projects. Significant investment into the R&D on the novel advanced accelerator techniques is required before one can evaluate opportunities for financially feasible, next generation energy frontier accelerators.